\documentclass[reprint,graphicx,amsmath,amssymb,]{revtex4-1}

\usepackage{graphicx}
\usepackage{dcolumn}
\usepackage{bm}
\usepackage{hyperref}
\usepackage{natbib}
\usepackage{acronym}
\usepackage{float}
\newcommand{\gr}{\Gamma_{\mathrm{Rb}}}
\newcommand{\gl}{\Gamma_{\mathrm{Li}}}

\begin{document}

\preprint{APS/123-QED}

\title{An Adaptable Dual Species Effusive Source and Zeeman Slower Design Demonstrated with Rb and Li}

\author{William Bowden}
\email{william.bowden@physics.ox.ac.uk}
\affiliation{The Department of Physics and Astronomy, University of British Columbia, Vancouver, Canada}

\author{Will Gunton}%
\affiliation{The Department of Physics and Astronomy, University of British Columbia, Vancouver, Canada}

\author{Mariusz Semczuk}%
\affiliation{The Department of Physics and Astronomy, University of British Columbia, Vancouver, Canada}

\author{Kahan Dare}%
\affiliation{The Department of Physics and Astronomy, University of British Columbia, Vancouver, Canada}

\author{Kirk W. Madison}%
\affiliation{The Department of Physics and Astronomy, University of British Columbia, Vancouver, Canada}

\date{\today}

\begin{abstract}
We present a dual-species effusive source and Zeeman slower designed to produce slow atomic beams of two elements with a large mass difference and with very different oven temperature requirements.  We demonstrate this design for the case of $^6$Li and $^{85}$Rb and achieve \ac{MOT} loading rates equivalent to that reported in prior work on dual species (Rb+Li) Zeeman slowers operating at the same oven temperatures.  Key design choices, including thermally separating the effusive sources and using a segmented coil design to enable computer control of the magnetic field profile, ensure that the apparatus can be easily modified to slow other atomic species.  By performing the final slowing using the quadrupole magnetic field of the MOT, we are able to shorten our Zeeman slower length making for a more compact system without compromising performance.  We outline the construction and analyze the emission properties of our effusive sources.  We also verify the performance of the source and slower, and we observe sequential loading rates of $8 \times 10^8$ atoms/s for a Rb oven temperature of $120\,^{\circ}$C and $1.5 \times 10^8$ atoms/s for a Li reservoir at $450\,^{\circ}$C, corresponding to reservoir lifetimes for continuous operation of 10 and 4 years respectively.

\end{abstract}

\pacs{Valid PACS appear here}
\maketitle

The ability to trap and cool multiple atomic species has garnered much interest within the cold atom community because the complex interactions within these systems give rise to a diverse range of physical phenomena.  Quantum degenerate Fermi-Fermi \cite{Dieckmann, Grimm}, Fermi-Bose \cite{InguscioAndRoati, BongsAndOspelkaus, HofstetterAndTitvinidze,PaulAndTung}, and Bose-Bose \cite{InguscioAndModugno, InguscioAndThalhammer, CapogrossoAndGuglielmino} gases allow for the study of novel states of matter which cannot be investigated in single species experiments. In particular, mixtures with large mass ratios, like those species presented here, are of interest for many body physics in the study of superfluidity \cite{GrimmAndJag}, spin impurities, and Effimov physics \cite{WeidemullerAndPires}. Unfortunately, large mass differences also results in practical challenges when trying to slow multiple species. 

Abundant samples of cold atoms are also a prerequisite for the formation of ultracold hetero-nuclear molecules \cite{YeAndNi,GrimmAndSpiegelhalder,ParkAndZwierlein} whose long range dipole-dipole interactions lead to exotic phases of matter and possible quantum information applications \cite{DeMille,ZollerAndMicheli,HazzardAndRey}. LiRb is an excellent candidate for such studies as it is predicted to have the second largest electric inherent dipole moment of the alkali dimers and when in the triplet state has the added advantage of a magnetic dipole moment \cite{Aymar}. Furthermore, for certain experiments, there are practical advantages to having the ability to trap multiple species. For example, one species with poor collisional properties (which limit the efficacy of evaporative cooling) can be cooled sympathetically via interactions with the other species \cite{Modugno09112001, GuptaAndVladyslav, SalomonAndSchreck, HadzibabicAndKetterle}. In other cases, one species can serve as an atomic detector to measure properties of the system, as has been demonstrated for thermodynamic measurements \cite{InguscioAndCatani, SalomonAndNascimbene}.

Creating large samples of cold atoms while still maintaining a good vacuum in the trapping region can be achieved by creating an atomic beam with an effusive oven separated from the trapping region by a differential pumping tube and by decelerating the beam before capture in a MOT using a Zeeman slower. Because the slowing of different species, especially those with very different masses, requires different magnetic field profiles, some multi-species experiments rely on a separate effusive source and Zeeman slower for each species.  However, to achieve a more compact setup, several realizations of multi-species Zeeman slowers have been developed (both dual \cite{KetterleAndStan, StamperKurnAndMarti} and triple species \cite{JulienneAndWille,WilleThesis}).  These approaches have dealt with the complications of dual species sources and Zeeman slowers in various ways.

In this work, we present and demonstrate a design that combines several key elements from previous work to achieve a compact, simple to fabricate dual--species effusive source and Zeeman slower adaptable to and effective for a wide variety of species combinations.  We experimentally demonstrate with Rb and Li that our design is well suited even for species combinations with large mass differences and with effusive sources requiring very different operating temperatures.  In particular, we observe loading rates of $8 \times 10^8$ $^{85}$Rb atoms/s for a reservoir temperature of $120\,^{\circ}$C and $1.5 \times 10^8$ $^6$Li  atoms/s with a reservoir temperature of $450\,^{\circ}$C, corresponding to reservoir lifetimes of four and 10 years respectively.  These Rb and Li loading rates are equivalent to those reported in prior work on multi-species Zeeman slowers operating at the same oven temperatures.

While our design achieves versatility and optimal performance for the sequential loading of the two species, in part, by including electronic switching between different magnetic field profiles as previously demonstrated by Paris-Mandoki {\it et al.} \cite{HackermullerAndParis}, our design differs substantially from that work in two key respects that make ours easier to build and more generally applicable to multi-species slowers.

The elements of our design include the following three key features.
\emph{(1) We create our atomic beam using a compact source composed of two thermally and physically separated effusive ovens with microtube array output nozzles for extended operational lifetimes.}  The separation of the beam output nozzles avoids the complications associated with a common mixing chamber such as back flow contamination of the reservoirs and chemical reactions within the mixing chamber \cite{KetterleAndStan, StamperKurnAndMarti} and allows for the independent and optimal temperature operation for each source.  Our design was inspired by the three species atomic beam created by Wille {\it et al.}; however, unlike their design which required highly specialized machining, our sources are made almost exclusively from standard vacuum parts \cite{JulienneAndWille,WilleThesis}.

\emph{(2) We generate our Zeeman slowing magnetic field profile with a segmented coil design} that enables computer control of the field profile thus allowing switching between the optimal operation of the slower for each species.  This feature also permits the adaptation of the slower to a new and different species without physical changes to the slower section.  Our slower coil consists of a set of eight independent solenoids and is very similar to the design shown to be very effective for a Cs+Li dual species slower created by Paris-Mandoki {\it et al.} \cite{HackermullerAndParis}.

\emph{(3) We mitigate the problem of beam blooming by disengaging the atoms from the Zeeman slowing light at the end of the slower using a magnetic field profile with a sharp drop created by a final disengagement coil and allowing the final stage of slowing to occur inside the MOT trapping region.}
Beam blooming occurs when the transverse velocity spread of the atomic beam (continually increasing because of transverse heating) becomes significant compared to the longitudinal velocity of the beam (which is decreasing during slowing).  If this occurs inside the slower or just beyond it, the atomic beam may diverge so much that a significant fraction of the atoms will not reach the MOT.  The growth of the transverse velocity spread because of the reemission of absorbed photons during slowing is proportional to the geometric mean of the recoil velocity and the velocity difference from slowing \cite{StamperKurnAndMarti}.  For this reason, beam blooming is a much more severe problem in light species (here the growth of the transverse velocity spread in Li is a factor of 8 larger than for Rb after slowing), and mitigating this effect has been approached in a variety of ways including adding transverse cooling inside the Zeeman slower \cite{JoffePritchard}, minimizing the distance (to 15~cm) between the end of the slower and the MOT \cite{StamperKurnAndMarti}, and matching the MOT quadrupole field to the Zeeman slowing field to achieve the final slowing stage inside the MOT region \cite{HackermullerAndParis}.  In order to avoid the additional expense and complications of a transverse cooling stage and to relax the constraint that the MOT be close to the end of the slower (required for the field matching slowing design of Paris-Mandoki {\it et al.} \cite{HackermullerAndParis}), we chose to simply disengage our atomic beam from the slower at a high velocity allowing the atoms to move a relatively long distance to the MOT region without loss from beam divergence and then to complete the final stage of slowing inside the MOT trapping region by using the MOT quadrupole field.  This feature was particularly important for our realization which involved the use of standard UHV parts including a long quartz to stainless transition connecting our quartz cell to our vacuum system.

Section 1 of this paper describes the design and construction of the Zeeman slower and effusive sources while highlighting some important considerations when designing multi-species cold atom experiments. Section 2 validates the performance of the system by loading MOTs of both species and discusses the results and their implications for slowing other atomic species.

\section{\label{sec:design}Design and Construction}

\subsection{\label{sec:sources}Effusive Sources}

Effusive sources are a common starting point for most laser cooling experiments and accurately predicting their emission properties is critical to ensuring proper performance.  Reloading the source after its depletion is a non-trivial task as it can potentially compromise the vacuum and often requires baking all or part of the system to achieve the desired base pressure. For this reason, it is desirable to maximize the lifetime of the effusive source without reducing the center line intensity of the atomic beam. One common approach is to use a recirculating or candlestick source where thermal gradients wick back and recollect atoms which are emitted off-axis \cite{ScholtenAndWalkiewicz}. The downside of such sources is their complexity in design and operation. An alternative is the use of arrays of large aspect ratio microtubes.  The microtubes collect the majority of atoms exiting off-axis and emit some fraction back into the reservoir without compromising the center line intensity of the atomic beam. This increases the longevity of the source without reducing loading rates. 

In the regime where the mean free path of atoms is much longer than the length of the tube, the rate at which atoms are emitted from the effusive source through an opening of area $A$ containing an atomic vapor of density $n$ with mean velocity $\bar{v}$ is

\begin{equation}
N=\frac{nA\kappa\bar{v}}{4} ,
\label{eq:totalFlux}
\end{equation}

\noindent where the parameter $\kappa$ is the Clausing factor and is derived from the geometry of the exit channel~\cite{clausing,beijer}. The design of such effusive sources for multi-species Zeeman slowers is complicated since they require collinear atomic beams. One approach is to connect multiple reservoirs to a common mixing chamber where the atoms exit via the same opening \cite{KetterleAndStan}. The steady state flux of the various species out of the effusive source is controlled by the rate they enter the mixing chamber. Care needs to be taken to prevent back flow between reservoirs and possible chemical reactions within the mixing chamber. Furthermore, the temperature of the mixing chamber must be kept warmer than the hottest reservoir to prevent condensation, which results in a higher mean velocity for atoms leaving the source than could otherwise be achieved with separate sources. This is especially problematic when the elements have vastly different vapor pressures at a given temperature, as it is the case when working with most alkali and alkali-earth mixtures. We elected to follow an alternative approach demonstrated by Wille \emph{et al.}~\cite{JulienneAndWille} of having separate sources with no mixing chamber to avoid such complications while still producing parallel and overlapping atomic beams by offsetting slightly their output ports.  This ensures our design is easily adapted to accommodate different atomic species.

To simplify the design over that presented by Wille \emph{et al.}, the entire Li source and the majority of the Rb source was made from off-the-shelf vacuum components. The Li source was made from a 1.33$''$ ConFlat nipple sealed on both sides with a ConFlat blank. A 4 mm $\times$ 2 mm oval was milled into the blank which connects the source to the apparatus into which approximately 60 microtubes are press fitted. The tubes are 1 cm in length with an inner (outer) diameter of 200 (300)~$\mu$m. Unlike other microtube designs which require a retaining plug or bar \cite{WeldAndSenaratne}, we found that press fitting the tubes held them sufficiently secure and well aligned. All gaskets within the Li oven are made from annealed nickel because copper cannot maintain a UHV seal as it is quickly corroded by the hot Li. During initial testing, copper gaskets were used and failed within a few hours to days depending on the operational temperature. Prior to loading the source, the Li (which was stored in a petroleum ether) was cleaned inside an argon filled glove bag by rinsing it in acetone and then cleaving off the outer oxidized layers.  Approximately 3 g of Li was loaded into the oven, which gives an estimate source lifetime of four years at a continual operational temperature of $450\,^{\circ}$C.  Common to most mircotube sources, the outlet is kept warmest to prevent clogging. However, the resulting temperature gradient leads to thermal wiking of the molten Li towards the hottest region (the outlet) as in the operation of a standard heat pipe. During testing, we observed molten Li being pulled through the microtube opening by capillary action which completely depleted the source. To prevent this, we lined the oven with a nichrome mesh to which Li preferentially adheres and helps to ensure it remains in the source. The mesh ensures that even with large temperature gradients, the liquid Li remains confined to the inside of the source and does not contact the microtubes.
  
\begin{figure}
\includegraphics[width=86mm]{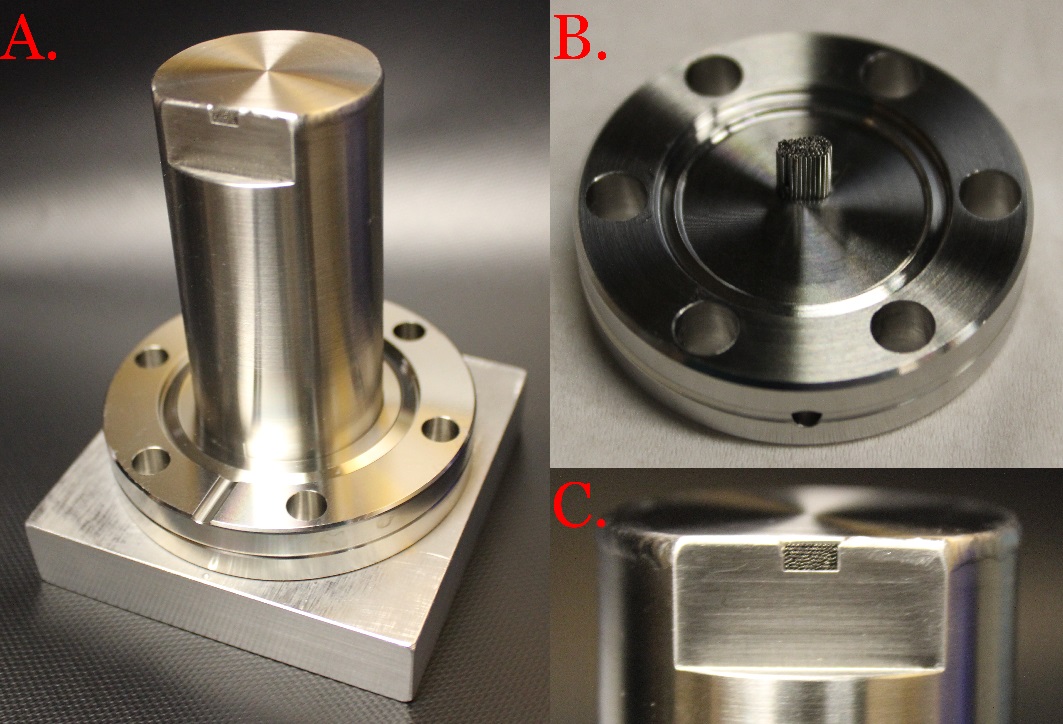}
\caption{\label{fig:ovenParts} Key components of the effusive sources: (A) Rb oven with microtube array, (B) Li oven cap with microtube array, (C) close up of Rb microtube array.}
\end{figure}

The Rb source is slightly more complex in design as the sample, which has a very high vapour pressure at elevated temperatures, must remain sealed in a glass ampule which is broken in-situ after the bakeout process. The ampule's base and top are held by two modified CF-blanks connected by a flexible bellows which is used to crack the ampule \cite{ScholtenAndBell}. In order for the Rb and Li atomic beams to be parallel and only slightly offset, the bellows is connected to a hollow stainless cylinder which extends into the vacuum system and ends directly below the Li exit channel. A 4~mm $\times~$2~mm opening is milled into a stainless steel cap, which is then TIG welded onto the top of the cylinder prior to being press fitted with microtubes. The key components of both sources are shown in \autoref{fig:ovenParts}.

\subsection{\label{sec:slower}Dual Species Zeeman Slower}

Zeeman slowers have been used extensively in ultracold atom experiments to produce beams of slow atoms as they are simple to construct, require relatively low power in the slowing lasers, and are robust against deviation in magnetic field or misalignment. Alternative approaches for loading \acsp{MOT} include transferring atoms from a 2D \ac{MOT} \cite{WeidemullerAndDieckmann, WalravenAndTiecke}, loading atoms directly from a background pressure produced by a dispenser \cite{YongAndDutta}, and capturing the unslowed atomic beam emitted by an oven \cite{Ladouceur}. The 2D \ac{MOT} approach has been successfully implemented for loading multiple species \cite{BongsAndOspelkaus2, SalomonAndRidinger}, but is significantly more complex and expensive than alternative designs (such as a Zeeman slower) primarily due the high power requirements for cooling. The dispenser approach is the simplest to implement, but only works for species with appreciable vapor pressure at room temperature such as Rb.  For elements such as Li where the vapor pressure is exceedingly low at room temperature, an oven producing an atomic beam with direct line of sight to the \ac{MOT} is required. Having dispensers or atomic ovens close to the trapping region can result in higher background pressures that limit the trap lifetime and thus the achievable atom number in the \ac{MOT}\cite{GoldwinAndJin,DieckmannAndTaglieber}. However, such an approach has proven successful in the production of degenerate gases of $^6$Li~\cite{Gunton}.

By placing the atomic oven further away from the trapping region, differential pumping techniques can be used to reduce the contribution of the background pressure in the trapping region due to the hot atomic sources. To offset the reduction in the trappable atom flux resulting from moving the oven away from the \ac{MOT}, a Zeeman slowing stage is used to slow fast moving atoms to a velocity which can be captured by the trap. 

Zeeman slowers are classified as one of three types depending on the polarization of the slowing light and magnetic field which in turn determines the atomic transition that is used for slowing. The classifications are $\sigma^+$ (decreasing magnetic field), $\sigma^-$ (increasing magnetic field), and spin-flip (magnetic field has zero crossing at some point in the slower) \cite{TruscottAndDedman}. The magnetic field profile for the $\sigma^+$ slower, which is presented in this work, is given by

\begin{equation}
B(z)=\frac{\hbar k v_c}{\mu}\sqrt{1-\frac{2za}{mv_c^2}}-\frac{\hbar\delta}{\mu}
\label{eq:slowerField}
\end{equation}

\noindent where $k$ is the wavevector of the slowing beam, $v_c$ is the capture velocity of the slower (i.e. the largest velocity class it can slow), $\mu$ is the magnetic moment of the transition, $a$ is the deceleration, and $\delta$ is the laser detuning.  We note that because of power broadening, the slowing of a particular velocity class can and may occur before it reaches the magnetic field at which it is exactly on resonance with the slowing light.  In addition, if the atoms temporarily fall out of the cycling transition and are repumped back into it (see discussion below), the slowing may occur just after the location in space where they are exactly on resonance with the slowing light.  The upper limit on deceleration, $a_{max}$, imposed by the finite scattering rate constrains the maximum magnetic field gradient which can be used for slowing to

\begin{equation}
\label{eq:maxGrad}
\left|\frac{dB(z)}{dz}\right| \ll \frac{\hbar k a_\mathrm{max}}{\mu v(z)},
\end{equation}

\noindent commonly referred to as the adiabatic slowing condition \cite{Napolitano}. In practice, the magnetic field is stretched spatially by a factor $\eta$ which relates the deceleration experienced by the atoms to $a_\mathrm{max}$ via $a=\eta a_\mathrm{max}$. If this gradient is exceeded, atoms will fall out of resonance with the slowing beam and will stop decelerating. In this work, we actually use this phenomenon to disengage the atoms from the slower to mitigate beam blooming and to ensure they are not stopped or turned around before reaching the \ac{MOT}. The unintended stopping and reversing of the atomic trajectories in the Zeeman slower is a primary consideration when optimizing its performance. 

The maximum gradient is typically different for various atomic species which limits the feasibility of simultaneous slowing. In general, the ratio

\begin{equation}
\frac{\eta_1}{\eta_2}=\frac{m_1\mu_1k_2\Gamma_2}{m_2\mu_2k_1\Gamma_1} ,
\end{equation} 

\noindent must be close to unity for efficient simultaneous slowing. For Li and Rb this ratio is 0.04 (predomintly due to the large mass difference). This greatly reduces the effectiveness of simultaneous slowing by using a fixed current or permanent magnets \cite{HillAndGill} design. Simultaneous slowing of the Li and Rb has been demonstrated, but it requires the magnetic field to be tailored such that specific regions along the slower cool different species \cite{StamperKurnAndMarti}. Alternatively, one could use multiple Zeeman slowers at the expense of reduced optical axis \cite{WeidemullerAndMosk}. Instead, we chose to dynamically switch the magnetic field profile to load the \ac{MOT} and \ac{ODT} sequentially. Although this results in slightly longer experimental cycle time, we can achieve much larger loading rates as we can independently tune the \ac{MOT} and Zeeman slower to the optimal loading parameters for both species. This sequential loading technique has been shown to be an effective approach to trapping Li and Rb in an ODT~\cite{PhysRevA.82.020701}.

\begin{figure}
\includegraphics[width=86mm]{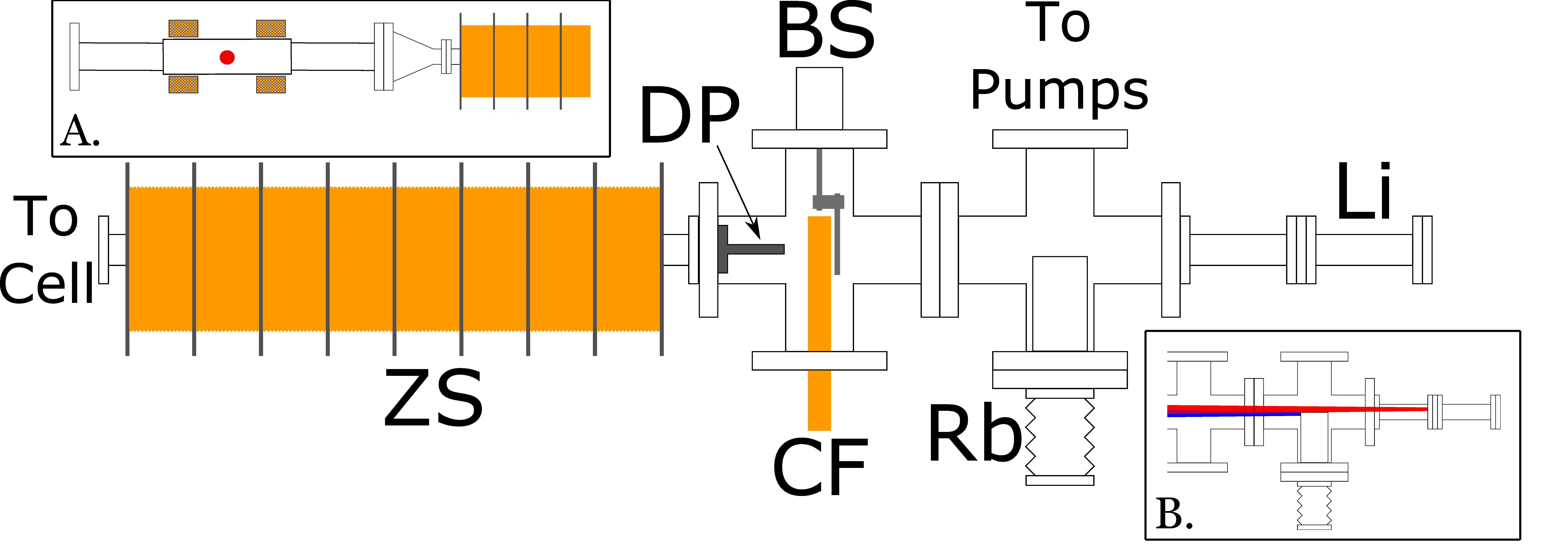}
\caption{\label{fig:setUp} Experimental apparatus showing the two effusive sources (Li, Rb), cold finger (CF), beam shutter (BS), differential pumping tube (DP), and Zeeman slower (ZS). The inset shows the trajectory of the two atomic beams for reference. Inset A shows the location of the \ac{MOT}, as indicated by the red circle, within the science section connected to the end of the slower.  Inset B shows co-propagating atomic beams from the effusive sources.}
\end{figure}

We elected to use a $\sigma^+$ design as the decreasing magnetic field has two main advantages. First, as a result of the atoms moving fastest in the large magnetic field region where the Zeeman effect almost cancels the Doppler shift, the required detuning of slowing beam is small and can be easily derived from our \ac{MOT} lasers using an acousto-optic modulator. Second, the decreasing magnetic field of the slower can be mated with the magnetic field from the quadrupole coils of the \ac{MOT} such that the slower length can be made significantly shorter as the atoms complete the final stage of slowing after entering the MOT trapping region.  Using the \ac{MOT} field to provide slowing has been used in our previous experimental apparatus \cite{Gunton} and in similar dual species slowers \cite{HackermullerAndParis}.

Critical to the success of our design, where the end of the slower and the MOT region are not in close proximity (required for smooth field matching between the end of the slower and the MOT region as demonstrated previously  \cite{HackermullerAndParis}), is our use of a coil at the end of the slower to produce a magnetic field with opposite polarity to disengage atoms from the slower by violating the adiabatic slowing condition. By varying the current, we can control the velocity at the point of disengagement to ensure the atoms are resonant with slowing laser when they reach the \ac{MOT} region.  As discussed earlier, using the \ac{MOT} field for slowing has the added benefit that the final stage of slowing is done close to the trapping region which helps to mitigate the blooming of the atomic beam. Blooming occurs because the slower reduces the velocity component parallel to the axis of the slower while heating the radial velocity distribution due to spontaneous re-emission of the absorbed light. This leads to an increasing divergence of the atomic beam as the atoms decelerate. The result of this is that the divergence angle of the beam upon leaving the slower (proportional to ratio of initial velocity to the final velocity) is continuously increased with more slowing and this counteracts the benefit of lengthening the slower in order to capture and slow atoms with even higher initial velocities. 

To produce the Zeeman slower field, we elected to use eight solenoids with computer controlled currents allowing for automated optimization of the magnetic field. Motivated by the fact that elongating the slower to increase the capture velocity leads both to increasing power dissipation and to diminishing returns due to beam divergence, we elected to build a relatively short slower measuring 24 cm which is easily air cooled using the metal fins which separate the coils. The slower is typically operated with less than a fifty percent duty cycle, but we find that the fins provide sufficient heat transfer to the air such that the slower can be operated continuously at the largest required currents with the hottest coil only reaching a temperature of $60\,^{\circ}$C. 

One drawback of the $\sigma^+$ design compared to a $\sigma^-$ and spin-flip slower, is that the slowing beam is much less detuned resulting in a larger radiation pressure exerted on the \ac{MOT}. To reduce the radiation pressure, we focus the slowing beam at the entrance of the slower such that the beam is large at the MOT and the divergence more closely matches that of the atomic beam.  This focussing has the added benefit that the beam curvature helps to provide some radial confinement.  In other work, adding a hole in the slowing beam has shown to effectively address this drawback \cite{PhysRevA.59.882}.

\begin{figure}
\includegraphics[width=86mm]{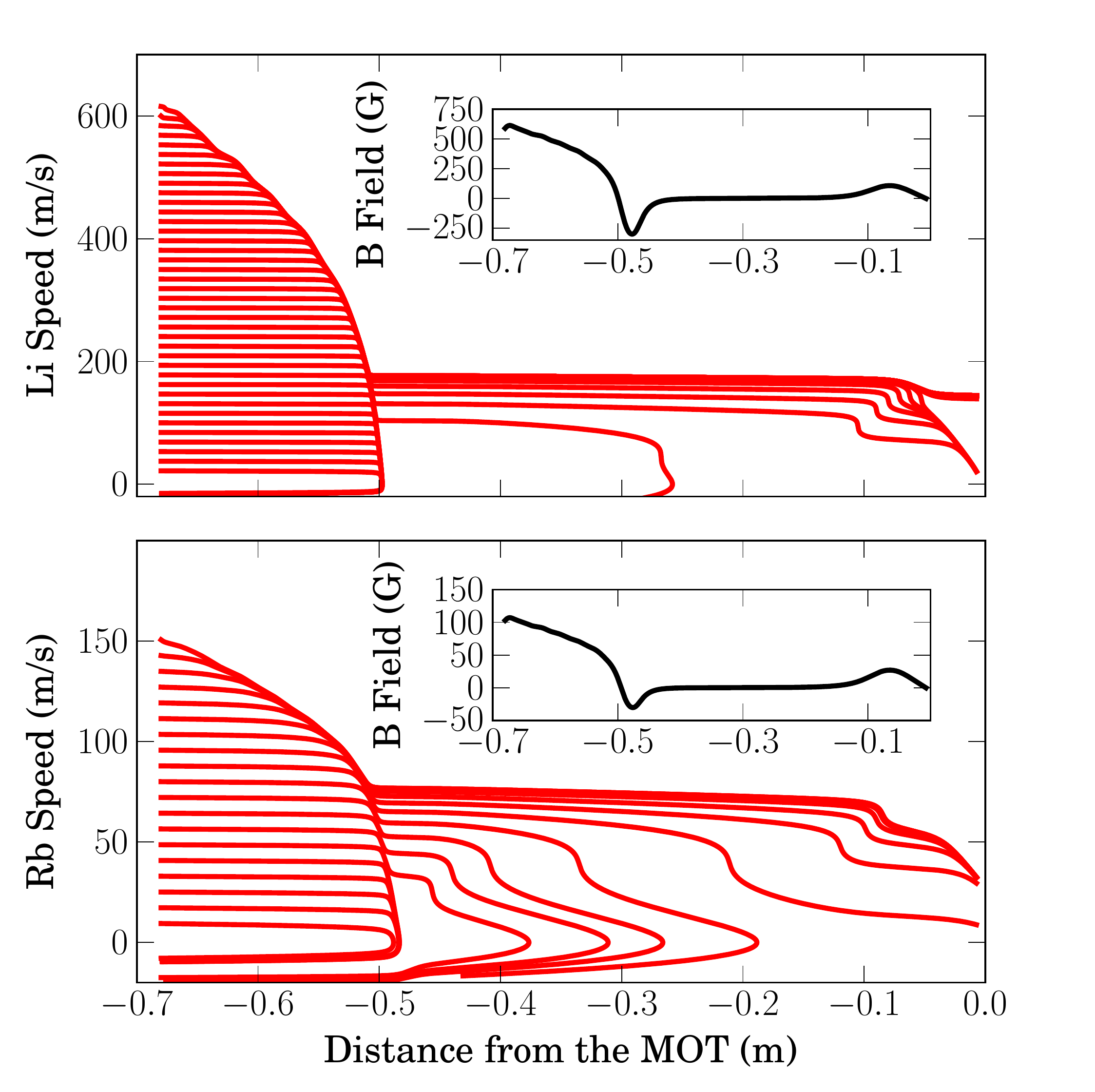}
\caption{\label{fig:phasespace}  Phase space trajectories for both species in the Zeeman slower and quadruple field of the \ac{MOT} coils. The magnetic field was chosen to illustrate how improper choice of the disengagement coil current can cause atoms to either pass directly through without further slowing (top panel) because they leave the slower moving too quickly, or turn around before reaching the trap (bottom panel) because the atoms leave the slower moving too slowly. Inset figures show the magnetic field produced by the \ac{MOT} coils and Zeeman slower}.
\end{figure}

Prior to constructing the slower, we simulated the atom trajectories as they decelerated within the slowing and trapping fields. The virtual slower adopts a two level model for the atom and solves the equation of motion subject to a positionally dependent radiation pressure. The result of the simulation is a bunching of atoms in phase space and can be used to estimate the proper current for the disengagement coil and slowing beam parameters. If the atoms leave the slower going too slowly they risk the possibility of being turned around, while if they are traveling too fast they will not be captured by the \ac{MOT}. The first effect is more likely for Rb as its exit velocity is lower than that of Li. These effects are illustrated in the phase space plots shown in \autoref{fig:phasespace}.

\subsection{\label{sec:vaccum}Vacuum System}

The Zeeman slower separates the science section of the experimental apparatus from the effusive source section. By decreasing the length of the slower, the isolation between the section is reduced requiring the addition of a differential pumping tube with an estimated hydrogen conductance of 1 L/s immediately before the slower. Both the source and science side are pumped by an Agilent VacIon Plus 20 Starcell ion pump and SAES CapaciTorr D 400-2 non-evaporatable getter. 

The Li effusive source is loaded with Li and baked separately at $500\,^\circ$C for 6 hours  in order to remove any remaining mineral oil or contaminants from the loading process before being back filled with argon and attached to the main apparatus. The entire apparatus is then baked for one week at $200\,^\circ$C. 

Long term exposure of the ion pumps to a significant background pressure of Rb generates filaments that, through field emission, produce large leakage currents.  This emission and heating limit the pumping performance by vaporizing the Rb and other materials originally stored in the pump.  It has been suggested that the lifetime of an ion pump exposed to a high vapor pressure of Rb can be prolonged by heating it continuously above its melting temperature \cite{Edwards}; however, we choose here to protect the pump by introducing a cold finger just after the Rb oven. Therefore, a copper feed through is placed inside the source section which is cooled externally using a TEC in order to condense Rb~\cite{KasevichAndAnderson}. 

Not shown in Fig.~\ref{fig:setUp} is an additional UHV cross to which is connected the Agilent VacIon Plus 20 Starcell ion pump and SAES CapaciTorr D 400-2 non-evaporatable getter.  Also on the cross is a window through which the Zeeman slowing light is introduced.
In order to minimize the coating of that window due to the Li atomic beam, we heat the window and we close the beam shutter (shown in Fig.~\ref{fig:setUp}) immediately after loading the MOT to minimize the time during which the Li beam is incident on the window.

\section{\label{sec:results}Results and Discussion}

\subsection{\label{sec:sourceresults}The Effusive Source}

Prior to assembling the entire apparatus, we measured the transverse velocity distribution of the emitted flux from the Li effusive source. To characterize the source, the emission was probed using a transverse laser which was scanned over the atomic D2 transition for the two ground hyperfine states. The flange containing the microtubes was heated to $450\,^{\circ}$C using metal band heaters while the opposite end was kept at approximately $15\,^{\circ}$C cooler. We developed a model~\cite{Bowden14} for the expected fluorescence signal based on the predicted angular and velocity distribution of the atoms leaving the effusive source. The expected angular distribution of a transparent channel of a given aspect ratio has been discussed in length in literature \cite{beijer, olander, hanes} and the velocity distribution, $F_{\mathrm{MB}}^{\mathrm{Beam}}(v)$, within the atomic beam emitted from a reservoir at a temperature $T$ is~\cite{ramsey}

\begin{equation}
\label{eq:velDistBeam}
F_{\mathrm{MB}}^{\mathrm{Beam}}(v)=\frac{m^2v^3}{2k^2T^2}\exp{\left[-\left(\frac{2kTv}{m}\right)^2\right]}.
\end{equation}

Based on the experimental data shown in Fig.~\ref{fig:ovenFlor}, the model estimates an average beam divergence of approximately $3\,^\circ$ which corresponds to a microtube aspect ratio of approximately 20; half the expected value of 40 given the microtube diameter (250~$\mu$m) and length (1~cm). A possible explanation could be a slight misalignment between microtubes or emission from gaps between tubes where they are not well packed. The total flux of atoms at this operational temperature was measured to be $9\times 10^{15}$ atoms/s while the predicted value using \autoref{eq:totalFlux} is $4\times 10^{15}$ atoms/s which is in reasonable agreement as it neglects any emission from the gaps between tubes or uncertainty in source temperature. The Rb source was not tested prior to installation into apparatus as it required breaking the ampule.

\begin{figure}
\includegraphics[width=86mm]{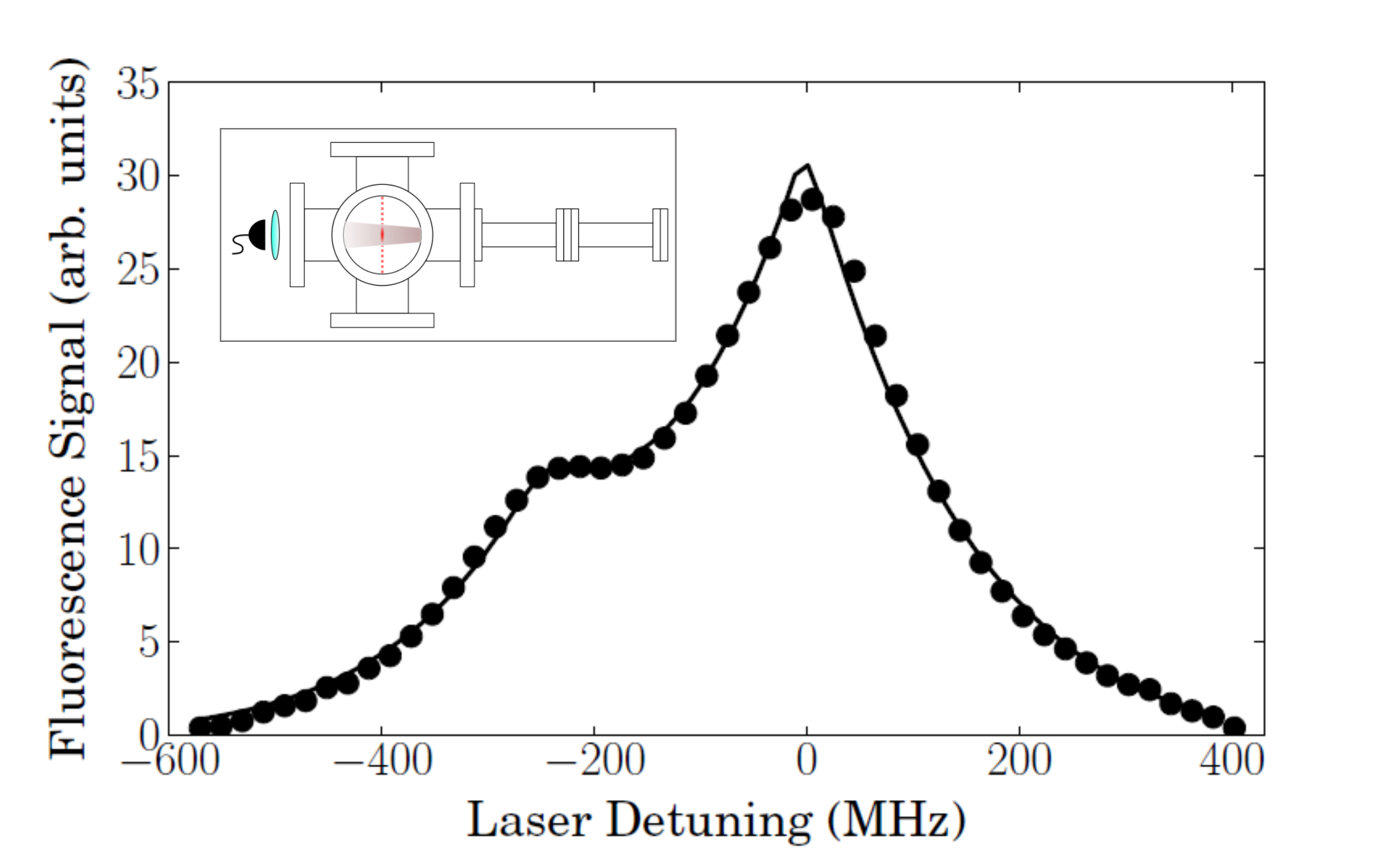}
\caption{\label{fig:ovenFlor}  Oven florescence (black circles) produced by the probe beam and the best fit of the numerical model (solid line, see text for details). The inset shows the experimental setup with the atomic beam going from right to left while probed by a transverse beam. In practice, the photodiode viewport is coated by Li during operation and it is best to place the detector along the other available axis orthogonal to the probe and atomic beams.}
\label{fig:ovenFlor}
\end{figure}

Characterization of the effusive source with respect to atom loading rate, \ac{MOT} lifetime, steady state atom number, and the lifetime of Li atoms loaded into an \ac{ODT} at various operation temperatures was performed with optimized settings for the Zeeman slower.  We note that the lifetime of atoms in the \ac{ODT} (given by the inverse single-particle loss rate) is a better and more relevant proxy for the quality of the vacuum than the MOT lifetime as there are additional loss mechanisms present in the \ac{MOT} other than just collisions with background gases.  These additional losses include light assisted collisions such as radiative escape and fine structure changing collisions, and they become more pronounced for larger \ac{MOT} atom numbers where the atomic density is higher.

The Li oven temperature was increased while the Rb source was kept at room temperature.  As the temperature increased, the loading rate and steady state atom number in the \ac{MOT} increased while the \ac{ODT} lifetime decreased steadily.  At the lowest operating temperature ($346\,^{\circ}$C), the Li \ac{MOT} lifetime is more than a factor of 2 longer than the \ac{ODT} lifetime.  This is expected given the \ac{MOT} trap depth is is more than three orders of magnitude larger than the \ac{ODT} \cite{MadisonAndVanDongen}.  However, as the source temperature is increased, the \ac{MOT} number and density grows, and the light assisted collisional losses in the \ac{MOT} become the dominant loss mechanism. The MOT lifetime is observed to then drop below the \ac{ODT} trap lifetime. Once the steady-state number is large enough, the \ac{MOT} grows in size with a constant density and the contribution to the single-particle loss rate (i.e.~lifetime) from light assisted losses becomes constant.  As the oven temperature is further increased, the \ac{ODT} lifetime is reduced due to increased collisions with background gases emitted by the oven. Because the \ac{ODT} depth is less than the \ac{MOT}, the lifetime reduction is larger for the \ac{ODT} than for the \ac{MOT}~\cite{VanDongen}.

To characterize the Rb slower, the Rb oven temperature was increased while the Li source was kept at a constant $356\,^{\circ}$C.  The loading rate and steady state atom number increase monotonically while the \ac{ODT} lifetime decreases.  For the Rb \ac{MOT}, the light assisted collisional losses are already dominant at the very lowest atom numbers (achieved at the lowest source temperature) and thus the lifetime of a Rb atom in the \ac{MOT} is always less than that of a Li atom in the \ac{ODT}. Note that for these measurements, the Rb \ac{MOT} was not present. Therefore, the Li lifetimes in the \ac{ODT} only represent the loss rate induced by collisions from atoms in the atomic beam and background gas and do not include any additional losses that might occur due to hetero-atomic light assisted collisions if the atomic clouds were well overlapped \cite{Ladouceur}.

\begin{figure}
\includegraphics[width=86mm]{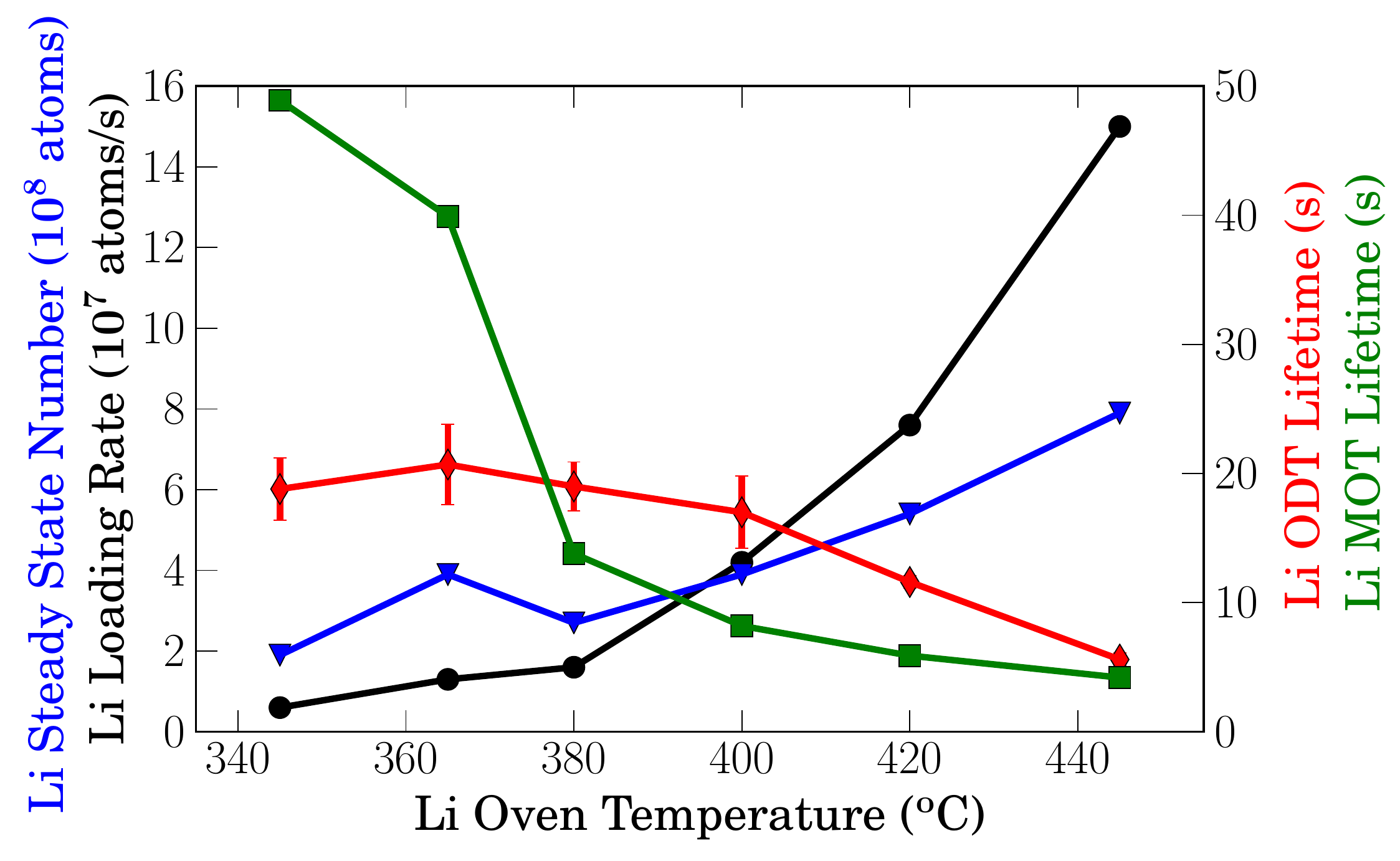}
\caption{\label{fig:liSourceTemp}  The effect of Li source temperature of Li loading rate (black dots), steady state atom number (blue trianges), \ac{MOT} lifetime (red diamonds), and \ac{ODT} lifetime (green squares). Due to density dependent loss mechanism within the \ac{MOT}, the lifetime of \ac{ODT} is a better estimate of the background pressure inside the science section of the apparatus.  Here, the Rb source is kept fixed at room temperature ($20\,^{\circ}$C, i.e., ``off"). Note the typical operating temperature of the Rb source is $100\,^{\circ}\mathrm{C}$}
\end{figure}

\begin{figure}
\includegraphics[width=86mm]{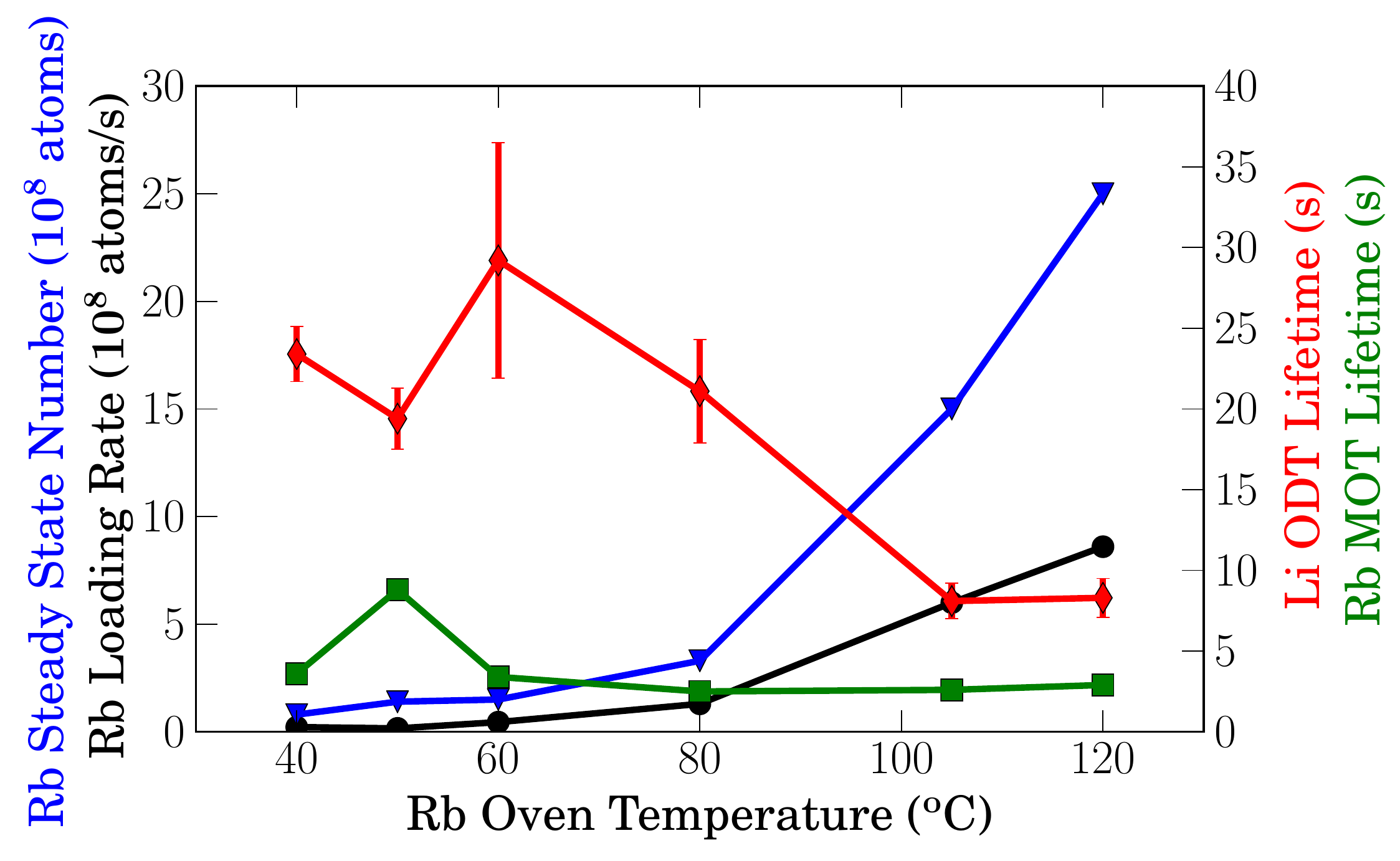}
\caption{\label{fig:rbSourceTempr}  The effect of Rb source temperature of Rb loading rate (black dots), steady state atom number (blue triangles), \ac{MOT} lifetime (red diamonds), and  lifetime for Li confined in a \ac{ODT} (green squares). For rubdium, light assisted collisional losses are dominant for even the lowest density traps resulting in lifetimes shorter than those observed for the Li \ac{ODT}. Here the Li source was held fixed at its standard operational temperature of $360\,^{\circ}$C.}
\end{figure}

\subsection{\label{sec:slowerresults}The Zeeman Slower}

Of interest for the design of the Zeeman slower is the effect of slower length and the intensity of the slowing light on loading rate.  In addition, the necessity of additional repump  light needed to optically pump atoms out of other ground states should be considered.  While we target with our $\sigma^+$ polarized light a closed (i.e.~cycling) transition for the slowing of both Rb and Li atoms, imperfect polarization can lead to off resonant excitation to excited state levels that can decay back to different ground states that are no longer in resonance with the slowing light and therefore are no longer decelerated effectively.
Empirically, we find that for the operation of the Rb Zeeman slower, an additional repump laser was required.  We find that such additional repumping light is not needed for Li, and we believe this is, in part, because the off resonant excitation to other levels is much more strongly suppressed than in Rb because of the much larger slowing magnetic fields used.  This is discussed more below.

For optimizing the \ac{MOT} loading rate, the expected detuning and field profile was set based on the virtual slower simulations, then each coil was scanned about its set point to optimize the loading rate. In all cases, the optimal setpoint was within a few percent of the predicted value. This process was repeated for increasing $\eta$ until a decrease in loading rate was observed.  Table \ref{tbl:MotParams} shows the relevant operational parameters used for testing. We were able to observe small \acp{MOT} for both species when the slowing beam is off. Activating the slowing beam and using the magnetic field produced by the quadrupole coils of the MOT as the sole Zeeman slower leads to a factor of 6-8 improvement in loading rate for both species. Activating the remaining Zeeman slower field further improved the loading by a factor of 5 and 12 for Li and Rb, respectively. 

\begin{table}[h]
\caption{Loading parameters for both \acsp{MOT}.}
\label{tbl:MotParams}
\centering
\begin{tabular}{l c c}
\hline
  & Rb & Li   \\ 
\hline
    Slowing Beam Pump Detuning (MHz)\footnote{The pump detuning for $^{85}$Rb is with respect to the $F=3\rightarrow F'=4$ D2 transition while for $^{6}$Li it is with respect to the $F=3/2\rightarrow F'=5/2$ transition at zero magnetic field.} & -85 & -76\\
	  Slowing Beam Pump Power (mW)  & 15 & 40 \\
    Slowing Beam Repump Detuning (MHz)\footnote{The repump detuning for $^{85}$Rb is with respect to the $F=2\rightarrow F'=3$ D2 transition while for $^{6}$Li it is with respect to the $F=1/2\rightarrow F'=5/2$ transition at zero magnetic field.} & 0 & --\\
	  Slowing Beam Repump Power (mW) & 12 & -- \\
		\acs{MOT} Pump Beam Detuning (MHz) & -15 & -45 \\
		\acs{MOT} Pump Beam Power (mW) \footnote{\label{beam_note}Beams have a radius of 9~mm, and the power is split between three retroflected arms of the \acs{MOT}} & 35 & 30 \\
		\acs{MOT} Repump Beam Detuning (MHz) & 0 & -40 \\
		\acs{MOT} Repump Beam Power (mW) \textsuperscript{\ref{beam_note}} & 10 & 40 \\
		\acs{MOT} Axial Gradient (G/cm) & 15.4 & 49 \\
\end{tabular}
\end{table}

To change the slower length, the number of coils activated was varied while monitoring the loading rate. To predict the loading rate, the flux of slowed atoms passing through the \ac{MOT} was calculated by integrating over the angular and velocity distribution of atoms leaving the effusive source. For each velocity, there is a critical exit angle for atoms leaving the effusive source above which they will miss the MOT after slowing. This angle is determined from basic kinematics given the constant deceleration along the slower and does not account for the added reduction in captured atoms due to transverse heating. This angle, along with the capture velocity of the slower, sets the limits of integration which in turn determine the flux that the MOT can capture. The scaling of loading rate with length is highly sensitive to the velocity group within the distribution being slowed. For Li, the velocity distribution is peaked at 1700~m/s, while the maximum capture velocity is approximately 600~m/s. As a result, the atoms slowed are from the the low velocity tail of the distribution. Directly integrating the Maxwell-Boltzmann distribution up to capture velocity for small velocities leads to $v^4$ scaling. Combining this with the $\sqrt{L}$ scaling for the capture velocity, one may expect to see quadratic scaling of the loading rate with initial increases in slower length if the divergence of the atomic beam is disregarded. In contrast, for Rb the distribution is peaked at much lower velocity of 300~m/s and atoms are predominantly captured from the linear region of distribution resulting in linear scaling of capture velocity with slower length using the same argument. Experimentally, we observe linear scaling of the loading rate for both species, and not quadratic for Li, which is in good agreement with our numerical simulation. The reduced loading rate for Li is a result of the larger divergence of the atomic beam upon exiting the slower, as compared to Rb, given its much higher initial velocity and lighter mass.

\begin{figure}
\includegraphics[width=86mm]{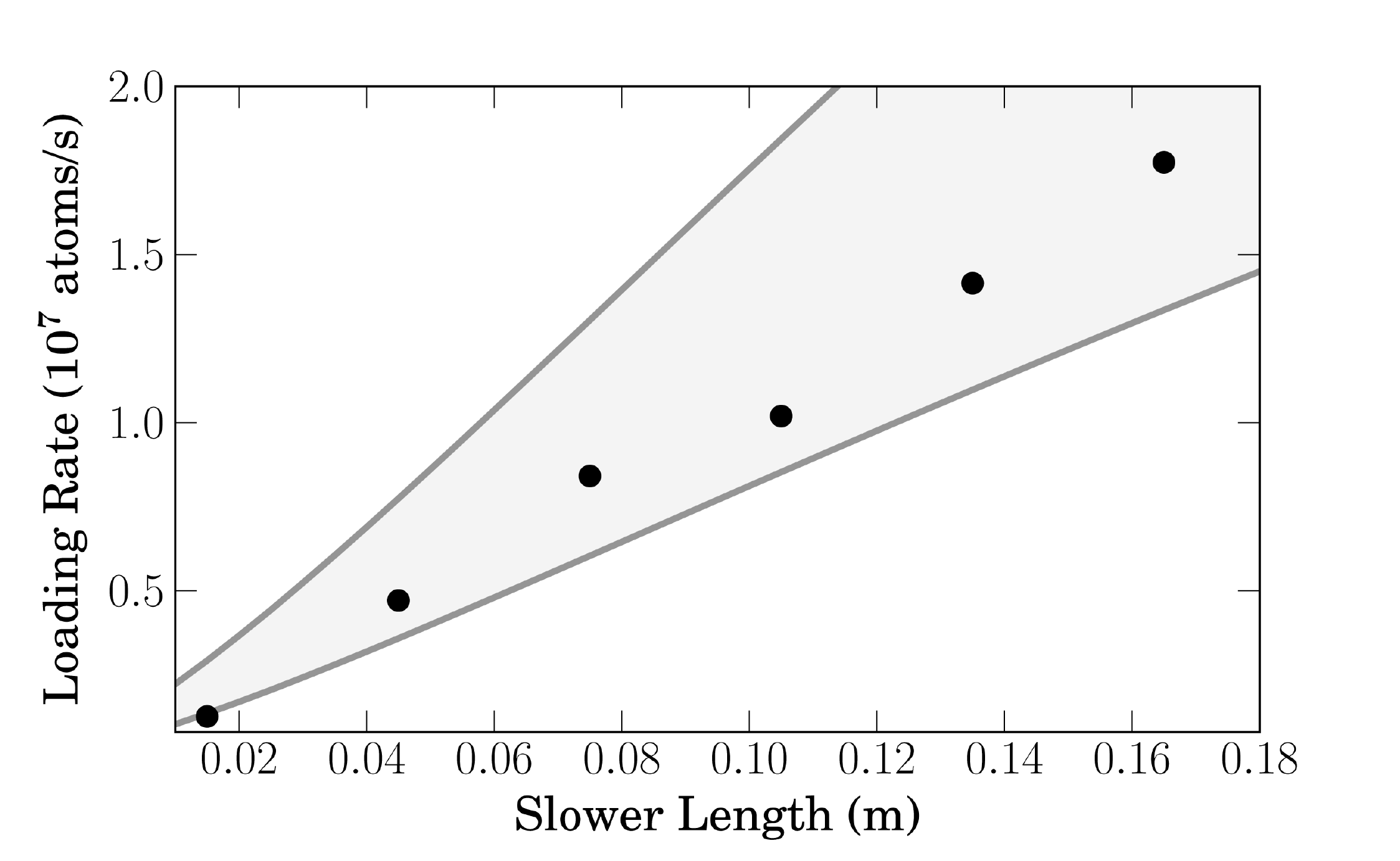}
\caption{\label{fig:loadingLi}  The \ac{MOT} loading rate for Li as function slower length at a source temperature of $360\,^{\circ}$C. The shaded region is the expected loading rate based on our model given a $\pm 5\,^{\circ}$C temperature uncertainty and $10\%$ uncertainty in microtube number.}
\end{figure}

\begin{figure}
\includegraphics[width=86mm]{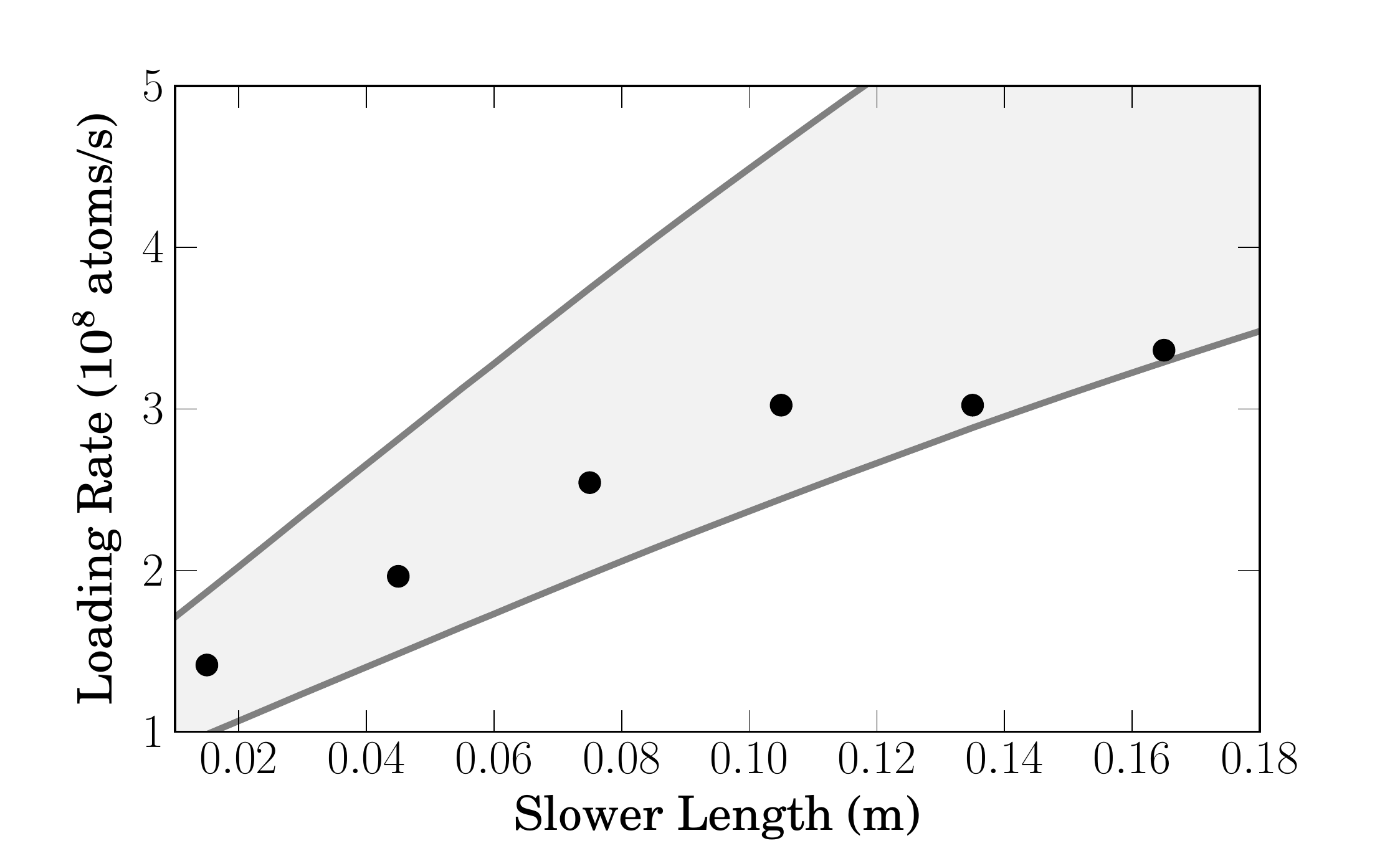}
\caption{\label{fig:loadingRb}  The \ac{MOT} loading rate for Rb as function slower length at a source temperature of $100\,^{\circ}$C. The shaded region is the expected loading rate based on our model given a $\pm 3\,^{\circ}$C temperature uncertainty and $10\%$ uncertainty in microtube number.}\end{figure}

It is challenging to accurately predict the loading rate for two reasons: 1) it is difficult to properly estimate the actual area of the effusive source outlet and account for microtube misalignment and the atomic emission from gaps and 2) it is difficult to know the exact source temperature to which the flux is exponentially sensitive.  In Figures \ref{fig:loadingLi} and \ref{fig:loadingRb}, we compare the measured and predicted loading rates on slower length, and we find that the model provides reasonable estimates for the expected flux given our uncertainty in the source temperature and a microtube array assembly ($10\%$ uncertainty in microtube number).

Finally the effect of the slowing beam intensity on loading rate was investigated.  For Li, we observed an improved loading rate with beam power while for Rb we saw decrease in loading rate at higher intensities.  We attribute the roll over of the Rb loading rate as resulting from atoms being stopped prior to reaching the \ac{MOT} at the higher slower intensities.  The Rb beam is more susceptible to this than the Li beam because of its much lower exit velocity and smaller slowing field.

\begin{figure}
\includegraphics[width=86mm]{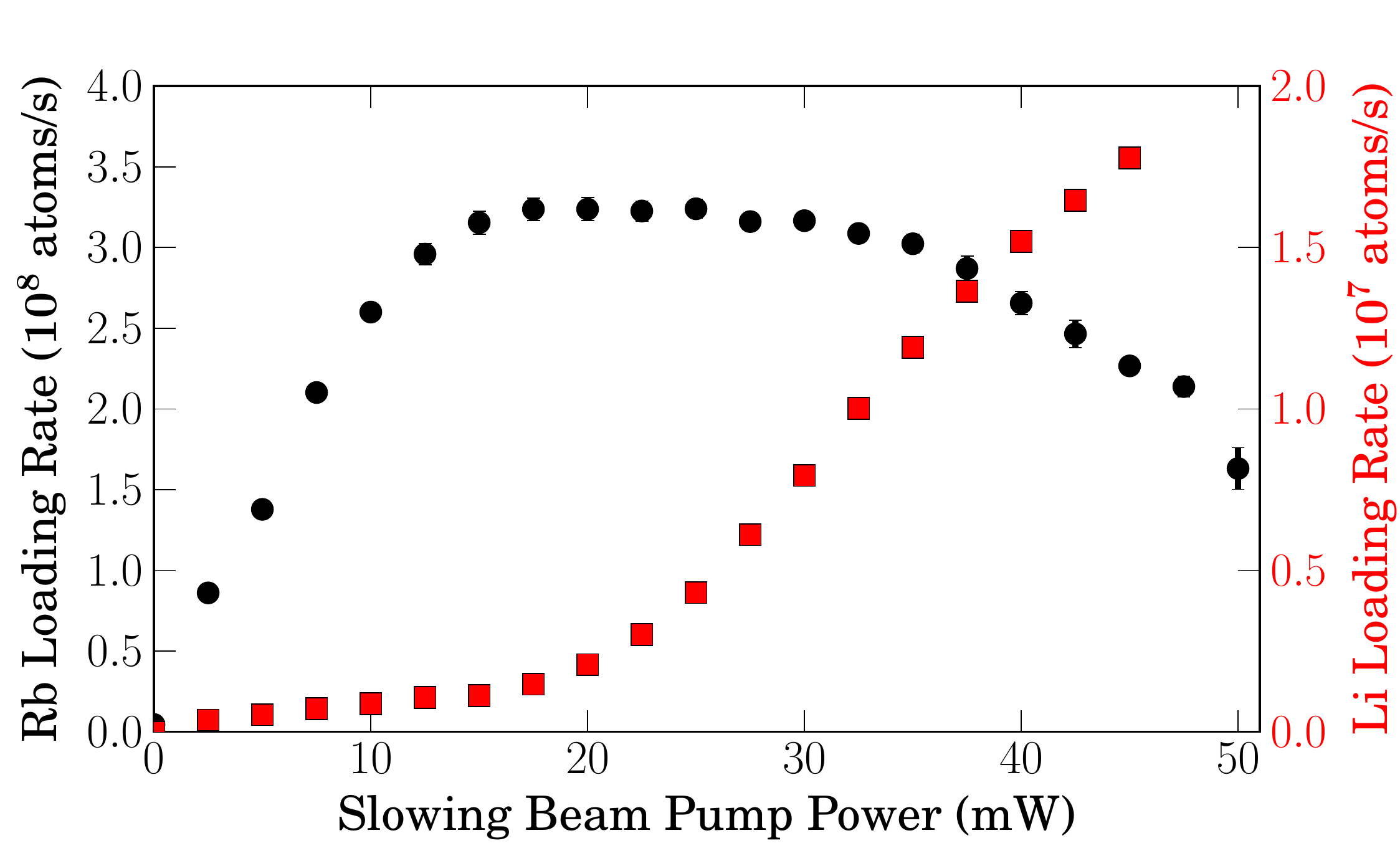}
\caption{\label{fig:slowingBeamIntensity}  The effect of slowing beam power on the \ac{MOT} loading rate for Li (red squares) and Rb (black dots). We attribute the roll over of the Rb loading rate as resulting from atoms being stopped prior to reaching the \ac{MOT} at the higher slower intensities.}
\end{figure}

\section{\label{sec:limitations} The role of repumping}

With this decreasing field slower, $\sigma^+$ polarized light is used and the magnetic field profile is optimized for atoms in the cycling (i.e.~closed) transition on the D2 line ($2^2 \mathrm{S}_{1/2} \rightarrow 2^2 \mathrm{P}_{3/2}$ for Li and  
$5^2 \mathrm{S}_{1/2} \rightarrow 5^2 \mathrm{P}_{3/2}$ for Rb) between the ground and excited stretched states.
These states correlate at zero magnetic field to 
$|F=3, m_F=+3\rangle$ and 
$|F'=4, m_F'=+4\rangle$ for $^{85}$Rb and
$|F=3/2, m_F=+3/2\rangle$ and 
$|F'=5/2, m_F'=+5/2\rangle$ for $^6$Li.
For strong magnetic fields, large enough that the energy shift due to the magnetic field is large compared to the hyperfine splitting, the hyperfine coupling between $J$ and $I$ is disrupted and $F$ is no longer a good quantum number.  In this case, the nuclear spin is decoupled from the electron and $m_I$ will not be changed by the absorption or emission of a photon for an electric dipole transition.  This limit is referred to as the hyperfine Paschen-Back regime and it occurs at fields above about 100~G for $^6$Li and above about a thousand Gauss for $^{85}$Rb.  For most of the initial slowing, Li is in this limit whereas Rb is not.  Thus the ground and excited stretched states in Li are more accurately labeled by their $m_J$ and $m_I$ quantum numbers:
$|m_J=1/2, m_I=+1\rangle$ and 
$|m_J'=3/2, m_I'=+1\rangle$.

The stretched-to-stretched state transitions are closed since a Rb(Li) atom that absorbs a photon and is excited into the $|F'=4, m_F'=+4\rangle$ ($|m_J'=3/2, m_I'=+1\rangle$) state can \emph{only} decay back to the stretched ground state $|F=3, m_F=+3\rangle$ ($|m_J=1/2, m_I=+1\rangle$).  However, leakage out of this cycling transition can occur if the polarization is not perfectly $\sigma^+$.  In this case, off resonant excitation to other excited states can occur, and subsequent decay out of these states may not necessarily bring the atom back to the stretched state but rather to other ground states.  In general, the optical transitions from these other ground states to the excited state manifold will not be in resonance with the slowing beam leading to a sharp drop in the photon scattering rate, and as a result the atoms will not be slowed properly as the adiabatic slowing condition will be broken.

In the case of Rb, when the polarization is not perfectly $\sigma^+$, off resonant excitation can occur from the stretched ground state to the $|F'=4, m_F'=+3 \; \, \mathrm{or} \, +2\rangle$ states (these transitions are quite likely since their frequencies at fields below 100~G are different from the stretched-to-stretched state transition by only a few tens of MHz corresponding to only a few natural linewidths, $\gr=2 \pi \times 5.7$~MHz) or to the $|F'=3, m_F'=+3 \; \, \mathrm{or} \, +2 \rangle$ states (the rate of these transitions is significantly lower than those to the $F'=4$ manifold because their resonant frequencies are different from the stretched-to-stretched state transition by a few hundred MHz at 100 G, but less than $50 \gr$).  Decay from the $F'=4$ states will return the atoms to the $F=3$ ground state, and at low magnetic fields (below 100 G) the transition frequencies from these states back to the $F'=4$ excited state are no more than a $5 \gr$ away from the slowing light frequency.  In this case, the scattering rate is still relatively high and atoms are optically pumped back into the stretched state before they have time to fall out of the slowing cycle by moving into a different region where they are off resonant with the slowing light.  In the case that off resonant excitation to the $F'=3$ manifold occurs, decay back to the $F=3$ or $F=2$ manifold is possible.  As discussed above, decay back to non-stretched states in the $F=3$ manifold does not lead to significant atom loss.  However, decay back to the $F=2$ manifold leads to a catastrophic drop in scattering rate from the slowing light since the transition frequency from the $F=2$ manifold to the $F'=3$ excited states is more than 3~GHz away ($500 \gr$) from the slowing light frequency.  Without light to move them back to the $F=3$ ground state manifold, the atoms are lost from the slowing.  

We note that for $\sigma^-$ (field decreasing) based Zeeman slowers where the opposite stretched states are targeted, the leakage out of the cycling transition due to imperfect polarization can be dramatically intensified.  This further necessitates the need for repumping atoms back to the cycling transition.  In $^{87}$Rb, this enhancement has been observed and occurs at a magnetic field of 120~G when the $|F'=3, m_F'=-3\rangle$ excited state crosses the $|F'=2, m_F'=-1\rangle$ excited state and the transition frequencies to this other excited state is exactly equal to the slowing light frequency \cite{GunterThesis}.  In $^{85}$Rb, this crossing occurs below 100~G and for $^6$Li it is irrelevant as it occurs below 10~G.

\begin{figure}
\includegraphics[width=86mm]{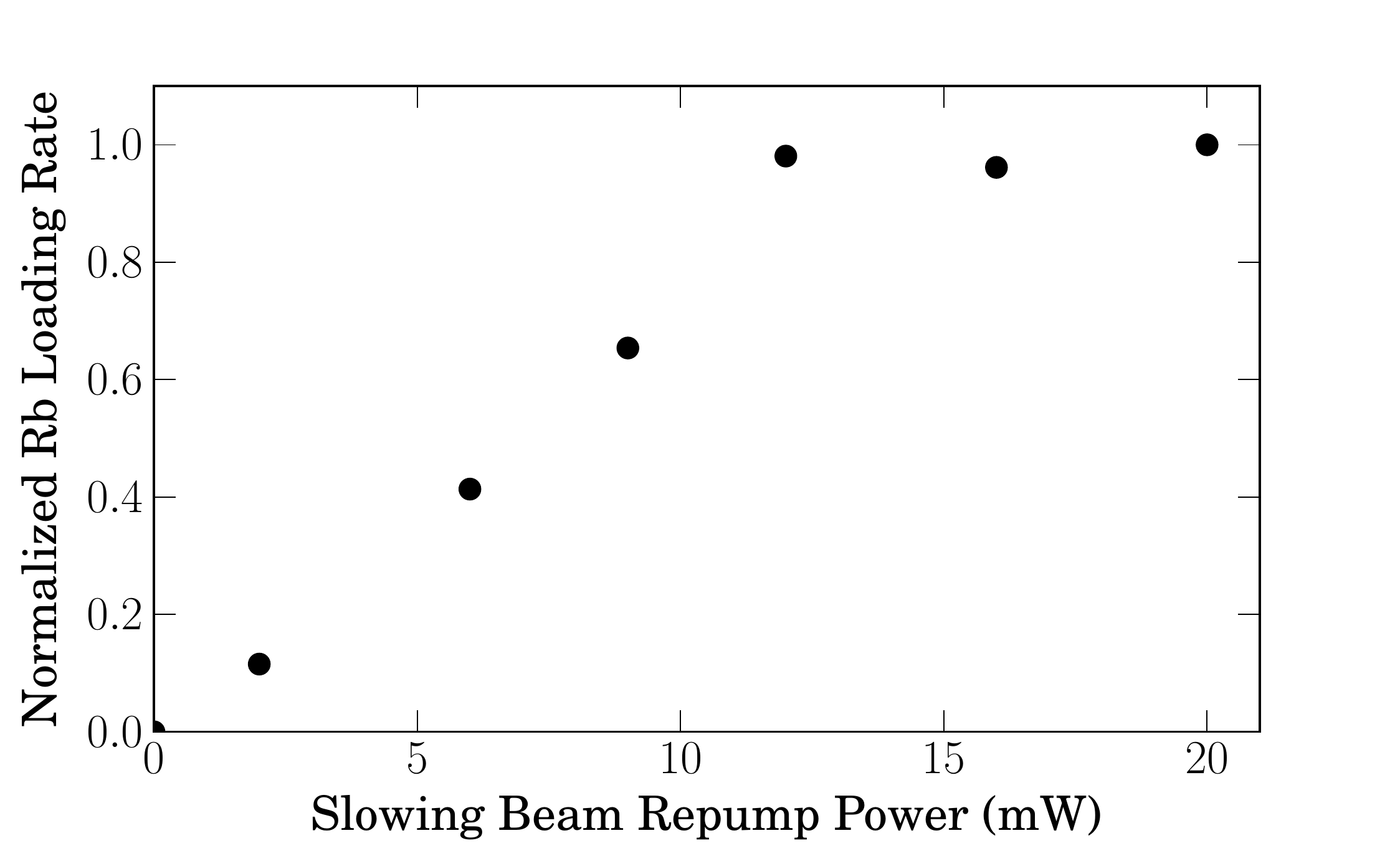}
\caption{\label{fig:loadingratevsrepump}  The effect of adding hyperfine repumping light to the slowing beam on the loading rate of the Rb MOT.  These data were taken using the parameters listed in Table~\ref{tbl:MotParams}, and the rate is normalized to the peak loading rate.}
\end{figure}

Figure \ref{fig:loadingratevsrepump} shows the impact of adding repumping light to the Zeeman slower beam on the loading rate of the Rb MOT.  In order to make this measurement, some of the repump light for the MOT was redirected into the slowing beam; however, since the slowing beam also traversed the MOT and had a similar size as the MOT beams at that location, redirecting light did not substantially change the total repump light provided to the MOT.  The frequency of the repump light added to the slowing beam was not frequency shifted as we the pump light.  Rather, the repump light was exactly on resonance for the $F=2 \rightarrow F'=3$ transition at zero magnetic field.  It is important to note that the Zeeman shift of this repump transition varies with the $m_F$ value in the ground state and it does not follow that of the stretched-to-stretched state transition, so there is no optimal detuning for the repump light.  Despite not being frequency shifted to compensate for the atomic beam's Doppler shift and the Zeeman shift, the repump light is nevertheless effective in returning the atoms to the cycling transition when they fall out of the stretched state.  In particular, we observe a strong dependence of the captured flux on repump power.  For a repump power of 12~mW we obtain the maximum in captured flux and additional light at this frequency does not further improve the loading rate.

In the case of lithium, we observed a large captured flux without any repump light added to the slowing beam.  While the same off resonant excitation mechanisms are present for lithium, there are several reasons why the leakage out of the stretched-to-stretched cycling transition is smaller than in Rb.  Firstly, Li is in the hyperfine Paschen-Back limit for the slowing within the Zeeman slower itself.  In this limit, the $m_I$ quantum number is unchanged by the absorption of a photon and therefore none of the other excited states (with $m_J'=3/2$) are excited by the slowing beam.  In fact, there is only one other excited state (other than the stretched excited state, $|m_J'=3/2, m_I'=+1\rangle$) to which the atom can go.  This is the $|m_J'=1/2, m_I'=+1\rangle$ state.  From this state, atoms can decay back to the initial stretched state or to the $|m_J=-1/2, m_I=+1\rangle$ ground state.  The transition frequency from this $m_J=-1/2$ ground state to the $m_J'=1/2$ excited state for is more than 1.4~GHz away from the slowing light frequency at the entrance of the Zeeman slower where the magnetic field is larger than 650~G.  Thus the scattering rate out of this state is very low and an atom that undergoes this off-resonant transition will be lost from the slowing cycle.
However, the frequency of the off resonant $|m_J=1/2, m_I=+1\rangle \rightarrow |m_J'=1/2, m_I'=+1\rangle$ transition is approximately 1~GHz (more than $150 \gl$) different from the slowing light frequency at 650~G, thus suppressing the rate of this off-resonant excitation by a factor of more than 2000 compared with the resonant scattering rate on the stretched-to-stretched transition (assuming a beam intensity of 40~mW/cm$^2$).  During the final stage of slowing where the magnetic field is much smaller, the resulting suppression of these off resonant transitions is smaller.  However, in our design, this final stage occurs inside the MOT region where repump light for the Li MOT may help to mitigate this effect.

We conclude by noting that we did not observe any improvement in the captured flux when the atomic beam was exposed to repump light just after the oven output.  In this case, the optical beam was perpendicular to the atomic beam and thus there was a negligible Doppler shift.  We performed this test for both Rb and Li to check if the captured flux could be increased by optically pumping the atoms into upper hyperfine levels (and thus increase the population of atoms in the stretched states) before they entered the Zeeman slowing region.  We did not observe an appreciable change in the captured flux.

\section{\label{sec:limitations} Complications of simultaneous, multi-species operation}

There are a number of complications inherent to all multi-species laser cooling experiments when simultaneous trapping of the species is performed.  For our application, the loading of the two species into an optical dipole trap, all of the complications arising from simultaneous slowing and simultaneous containment of both species in the MOTs can be avoided.  In short, the sequential loading of the MOT and transfer of the atoms into the ODT is very effective, and this approach was used successfully in prior measurements of the Feshbach resonances in the Li+Rb mixture \cite{PhysRevA.82.020701}.  In that work, the ODT and MOTs were not overlapped until the moment of transfer, and the only atom loss incurred for the first species transferred was due to background collisions during the loading of the second species.  For high loading rates (less than a second) and long trap lifetimes (more than 10 seconds) the loss from waiting for the other species to load is on the order of 10\% or less.

However, there are applications for which simultaneous loading and storage in the MOT may be necessary.  Two examples are photoassociation studies with the atoms held in the MOT and the transfer of the laser cooled mixture to a magnetic trap for which sequential transfer made not be possible.

For simultaneous slowing and trapping of multiple species, the first and most obvious complication is that the optimal Zeeman slower field profile and the optimal MOT magnetic field gradient may be different to the point of being incompatible for different species.  For example, here we find empirically that the optimal MOT axial field gradients for Rb and Li are 15.4 and 49 G/cm respectively, differing by more than a factor of 3, and the optimal initial field of the Zeeman slower for Li and Rb differs by a factor of 6 due to the different adiabatic conditions. Consistent with prior observations, the Li MOT performance at the detunings and power listed in Table~\ref{tbl:MotParams} showed dramatic improvements at high field gradients (above 30 G/cm) and the Rb atom number in the MOT is dramatically suppressed when operated at field gradients above about 28 G/cm \cite{Ladouceur}.  These differences arise from an interplay between the homonuclear light assisted collisional loss rates and the trap-depth recapture probability of the products of inelastic collisions (see the work of Ladouceur \emph{et al.} and references therein \cite{Ladouceur}).  Of course, if simultaneous operation of the MOTs is required, a new field gradient and MOT parameters can be found that are acceptable yet non--optimal for each species.  

An additional complication arises in simultaneous operation of the MOTs when the atom clouds overlap.  In that case, heteronuclear light-assisted losses can occur further suppress the steady state atom numbers in the MOT.  Prior work has shown that, under certain conditions, the Li atom number is more dramatically reduced than the Rb atom number presumably due to differences in the MOT trap depths and the re-capture probabilities of the collision partners accelerated by these these binary inelastic collisions \cite{Ladouceur}.  In the work presented here, because we use a retro-reflection configuration for the MOT beams and the return beams are less intense, the radiation pressure is imbalanced and the Rb and Li MOTs are observed to be offset from one another by a distance more than their diameters.  This offset is even more pronounced due to the different offsets produced by the Zeeman slowing beams.  Thus we do not observe significant light assisted losses when both species are present in the MOT.

\section{\label{sec:conclusion}Conclusion}

We have presented a design for a multi-species effusive source and slower which is applicable to the slowing of a wide variety of species, even those with large mass differences, and demonstrated the design with Li and Rb.  Our choice of slower length was made to balance the marginal returns on performance (as there is a linear scaling of flux with slower length) with the goal of creating a design that is simple (for example, compact and without the need for water cooling) and easily adaptable to other atomic species.  By utilizing the quadruple magnetic field for the \ac{MOT} as a secondary slowing field, we were able to further shorten the length of the slower.  While it appears that a longer slower would improve Li loading rates, we achieved comparable or higher loading rates to prior work at similar oven operating temperatures.  In particular, we observe loading rates of $8 \times 10^8$ atoms/s for a Rb oven temperature of $120\,^{\circ}$C and $1.5 \times 10^8$ atoms/s for a Li reservoir at $450\,^{\circ}$C, corresponding to reservoir lifetimes for continuous operation of 10 and four years respectively.  Based on simulations, we believe that for Li, the linear scaling was a result of divergence of the atomic beam, while for Rb it was due to the peaking of the thermal distribution of velocities leaving the source.

\section{\label{sec:ack}Acknowledgments }

The authors also acknowledge financial support from the Natural Sciences and Engineering Research Council of Canada (NSERC / CRSNG), and the Canadian Foundation for Innovation (CFI).  W.B.~would also like to also acknowledge NSERC support through the Canada Graduate Scholarships-Master's program.  This work was done under the auspices of the Center for Research on Ultra-Cold Systems (CRUCS). Finally, the authors would also like to thank Florian Schreck for his generosity with superb technical advice and the microtubes used for both sources.

\bibliography{myBib}

\begin{acronym}[abc]
  \acro{MOT}{magneto-optical trap} 
	\acro{ODT}{optical dipole trap} 
\end{acronym}

\end{document}